%% file: main-tvcg.tex
\documentclass[journal, preprint]{vgtc}
\usepackage{etoolbox}
\patchcmd{\thebibliography}{\clubpenalty4000}{\clubpenalty10000}{}{}
\patchcmd{\thebibliography}{\widowpenalty4000}{\clubpenalty10000}{}{}

\ifpdf%                                % if we use pdflatex
  \pdfoutput=1\relax                   % create PDFs from pdfLaTeX
  \usepackage{graphicx}                % allow us to embed graphics files
  \DeclareGraphicsExtensions{.pdf,.png,.jpg,.jpeg} % for pdflatex we expect .pdf, .png, or .jpg files
\else%                                 % else we use pure latex
  \usepackage{graphicx}                % allow us to embed graphics files
  \DeclareGraphicsExtensions{.eps}     % for pure latex we expect eps files
\fi%

\graphicspath{{figures/}{pictures/}{images/}{./}} % where to search for the images

\usepackage{microtype}                 % use micro-typography (slightly more compact, better to read)
\PassOptionsToPackage{warn}{textcomp}  % to address font issues with \textrightarrow
\usepackage{textcomp}                  % use better special symbols
\usepackage{mathptmx}                  % use matching math font
\usepackage{times}                     % we use Times as the main font
\usepackage{cite}                      % needed to automatically sort the references
\usepackage{tabu}                      % only used for the table example
\usepackage{booktabs}                  % only used for the table example
\usepackage{amsmath}
\usepackage{amsfonts}
\usepackage{mathtools}

\usepackage{algorithm}
\usepackage{algpseudocode}

\usepackage{cite}                      % needed to automatically sort the references
\usepackage{tabu}                      % only used for the table example
\usepackage{booktabs}                  % only used for the table example
\usepackage{multirow}

\input{macros.tex}

\usepackage{multicol}

\usepackage{xspace}
\newcommand{\etal}{et al.\@\xspace} % Prints ``et al.'' with proper spacing
\newcommand{\ie}{i.e.\@\xspace} % Prints ``i.e.'' with proper spacing
\newcommand{\commentout}[1]{}

\usepackage{color}
\usepackage{savesym} \savesymbol{spacing} % to avoid conflict with setspace
\usepackage{setspace}
\definecolor{Orange}{rgb}{1,0.5,0}

\definecolor{DarkGreen}{rgb}{0,0.5,0}
\definecolor{Purple}{rgb}{0.7,0,0.7}
\definecolor{Blue}{rgb}{0.2,0.2,0.8}
\definecolor{Red}{rgb}{1.0,0.0,0.0}
\definecolor{Brown}{rgb}{0.7,0.4,0.1}

\usepackage[whole]{bxcjkjatype}
\onlineid{1879}

\vgtccategory{Research}

\preprinttext{To appear in an IEEE VGTC sponsored conference.}

\title{ChromaGazer: Unobtrusive Visual Modulation using\\Imperceptible Color Vibration for Visual Guidance}

\author{
    Rinto Tosa\thanks{e-mail: rinto\_tosa@pml.slis.tsukuba.ac.jp}\\{\scriptsize \centering University of Tsukuba}
    \and Shingo Hattori\thanks{e-mail: s.hattori@cluster.mu}\\{\scriptsize \centering Cluster Metaverse Lab}\\{\scriptsize \centering University of Tsukuba}%
    \and \authororcid{Yuichi Hiroi}{0000-0001-8567-6947}\thanks{e-mail: y.hiroi@cluster.mu}\\ {\scriptsize \centering Cluster Metaverse Lab}
    \and \authororcid{Yuta Itoh}{0000-0002-5901-797X}\thanks{e-mail: yuta.itoh@iii.u-tokyo.ac.jp}\\{\scriptsize \centering The University of Tokyo}
    \and \authororcid{Takefumi Hiraki}{0000-0002-5767-3607}\thanks{e-mail: t.hiraki@cluster.mu}\\{\scriptsize \centering Cluster Metaverse Lab}\\{\scriptsize \centering University of Tsukuba}%
}
\teaser{
    \includegraphics[width=\linewidth]{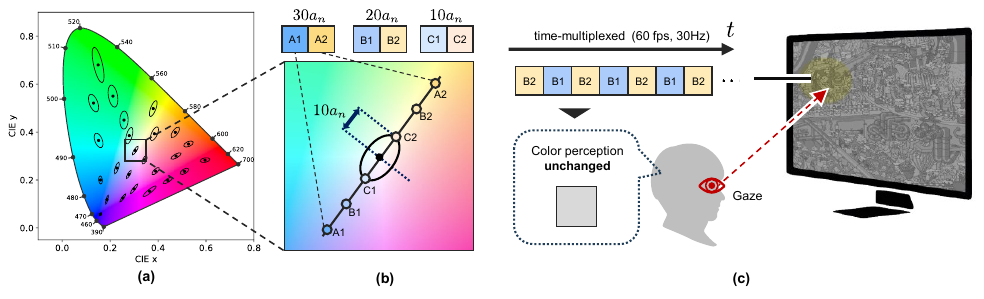}
        \caption{\label{fig:teaser} 
        Overview of our unobtrusive visual guidance (VG) using color vibrations. (a) MacAdam ellipse~\cite{MacAdam1942-sv} contains colors that are indistinguishable from the average human eye when juxtaposed with the color at the center of the ellipse on $xy$ chromaticity diagram. For illustration, the ellipses are shown at 10 times their actual size. (b) We select color pairs for color vibration by multiplying the major axis $a_n$ of each MacAdam ellipse by the amplitude $r$. We empirically determine the range of $r$ suitable for unobtrusive VG. (c) By modulating regions of interest in an image with the $r$ determined in (b), we implement VG without significantly altering the perceptual appearance of the image.
    }
}

\abstract{
Visual guidance (VG) is critical for directing user attention in virtual and augmented reality applications. However, conventional methods using explicit visual annotations can obstruct visibility and increase cognitive load. 
To address this, we propose an unobtrusive VG technique based on color vibration, a phenomenon in which rapidly alternating colors at frequencies above 25 Hz are perceived as a single intermediate color. 
We hypothesize that an intermediate perceptual state exists between complete color fusion and perceptual flicker, where colors appear subtly different from a uniform color without conscious perception of flicker. To investigate this, we conducted two experiments. First, we determined the thresholds between complete fusion, the intermediate state, and perceptual flicker by varying the amplitude of color vibration pairs in a user study. Second, we applied these threshold parameters to modulate regions in natural images and evaluated their effectiveness in guiding users' gaze using eye-tracking data. Our results show that color vibration can subtly guide gaze while minimizing cognitive load, providing a novel approach for unobtrusive VG in VR and AR applications.
}


\begin{document}

\firstsection{Introduction}

\maketitle

Visual guidance (VG) is crucial for directing users' attention to specific areas of interest by overlaying information on virtual or real scenes. Beyond traditional applications in graphics and web design~\cite{Pang2016Direct}, VG has gained prominence in virtual reality (VR) and augmented reality (AR) environments, such as virtual training~\cite{Wang2022Attention}, exploration support~\cite{McNamara2008Search}, and memory support~\cite{Eduardo2011Direct}. However, traditional VG methods that use explicit visual annotations such as arrows or circles~\cite{Reif2009pickby, Renner2020ARglass, Benjamin2018sar} can increase cognitive load and obscure target objects.

To address these challenges, researchers have explored \textit{unobtrusive} VG techniques that minimize visual distraction while preserving scene context. One approach is to manipulate visual saliency~\cite{koch1987attention}, the perceptual quality that makes certain elements stand out and attract attention. Previous studies have modulated various aspects of visual saliency, such as color and contrast~\cite{azuma2018gaze, Nguyen2013reattention, Mateescu2014attention, Sutton2022}, or introduced flicker~\cite{Bailey2009subtle, McNamara2008Search, Sutton2024flicker}, to direct users' gaze to regions of interest. These methods aim to enhance the salience of target objects without adding extraneous graphical elements.
However, most saliency-based techniques inevitably alter the appearance of the scene, potentially altering the meaning of the content, causing important information to be missed, and reducing consistency with the real world in AR applications. 

To overcome these limitations, we introduce ChromaGazer, a VG method that uses imperceptible color vibration to guide the user's gaze without altering the visual appearance of the content.
Imperceptible color vibration occurs when two colors of the same luminance alternate at frequencies above the critical color fusion frequency (CCFF)~\cite{Lange1958ccff}, approximately 25~Hz. At these frequencies, the human visual system perceives the alternating colors as a single fused color. We hypothesize that by carefully selecting color pairs and adjusting their vibration amplitude, it is possible to create an intermediate perceptual state that subtly attracts users' attention to target regions without them consciously perceiving changes in the object's appearance. We define this intermediate state as one in which the target appears slightly different from a monochromatic color, but the flicker is not overtly perceived.

To assess the feasibility of using color vibration for VG, we conducted two experiments. In the first experiment, we systematically varied the vibration amplitude of color pairs based on the MacAdam ellipse~\cite{MacAdam1942-sv} and asked participants to identify thresholds between three perceptual states: explicit flicker, intermediate, and fused colors. By determining these thresholds, we aimed to identify optimal parameters for producing the desired intermediate state. In the second experiment, we applied color vibrations with the parameters identified in the first experiment to specific regions within natural images. We then evaluated their effectiveness in directing participants' gaze using eye-tracking data. The results showed that our method successfully redirected users' gaze without their conscious awareness.
% \note{WILL REVISE according to the experiment and results.}

Our main contributions include:
\begin{itemize}[leftmargin=*]
\setlength{\parskip}{0.1cm}
    \item We propose ChromaGazer, a VG technique that uses imperceptible color vibrations to guide the user's gaze without altering the appearance of the scene.
    \item We identify the optimal parameters for creating an intermediate state that balances the attention and visual integrity of content through a user study.
    \item We demonstrate the effectiveness of our color-vibration-based VG method in guiding users' gaze to target regions within natural images, as measured by eye-tracking data.
    \item We explore the design considerations of our unobtrusive VG approach using color vibration and suggest avenues for future research.
\end{itemize}

\section{Related Work}\label{sec:related}

\subsection{Visual Guidance}
Visual guidance has been extensively studied due to its wide range of applications, such as guiding users through digital information~\cite{Pang2016Direct, McNamara2008Search}, assisting in industrial tasks~\cite{Schwerdtfeger2008order}, and supporting training~\cite{Eduardo2011Direct, yoshimura2019visualcue, Wang2022Attention}. The control of visual attention involves two mechanisms: top-down and bottom-up attention~\cite{Posner1980attention}. Top-down attention is controlled by internal factors such as intentions, goals, and knowledge, and is typically attracted by methods such as arrows, icons, or textual instructions. Conversely, bottom-up attention is controlled by external factors, such as the characteristics of visual stimuli such as flickering lights, bright colors, and motion.

Traditionally, most visual guidance techniques have relied on approaches that attract top-down attention, such as outlines like circular or rectangular frames~\cite{Jo2015eyebookmark, Kern2010gazemarks, Waldner2014attractive} or arrows~\cite{schmitz2020directing, wallgrun2020comparison, yoshimura2019visualcue}. On the other hand, methods that use bottom-up attention for visual guidance have also been the subject of longstanding research. These approaches manipulate the visual saliency of specific elements within a visual scene, making them stand out from other elements. Various parameters for adjusting scene saliency have been investigated, with most methods falling into two categories: color adjustment~\cite{azuma2018gaze, Nguyen2013reattention, Mateescu2014attention, Sutton2022} or flicker control~\cite{Bailey2009subtle, McNamara2008Search, Sutton2024flicker}. In addition, control techniques based on saliency maps~\cite{koch1987attention}, which quantify visual saliency using insights from neuroscience, have been proposed. For example, Kokui~\etal~\cite{kokui2013color} applied color shifts based on saliency maps, while other studies have focused on modulating spatial frequency or texture power maps.
Furthermore, researchers have explored combining multiple parameters that influence saliency to achieve more effective gaze guidance. Examples include the simultaneous modulation of color and brightness~\cite{hagiwara2011saliency, shi2014videosaliency}, the introduction of subtle blur effects to modulate visual saliency~\cite{hata2016visual}, and the application of genetic algorithms~\cite{Pal2017genetic}. 

Recently, Suzuki~\etal proposed an approach that combines various parameters that affect visual saliency, such as blur, brightness, saturation, and contrast~\cite{suzuki2017saliency}. 
Also, Sutton~\etal proposed a method of gaze guidance using a see-through optical head-mounted display to modulate multiple visual salience parameters of a real see-through view~\cite{Sutton2022}.
However, most existing saliency manipulation methods inevitably alter the appearance of the content, which can be problematic. This study investigates a novel approach that exploits the phenomenon of imperceptible color vibration to control visual saliency and guide gaze without altering the appearance of the content.

\subsection{Imperceptible Color Vibration}
Imperceptible color vibration exploits the human visual property that observers perceive an intermediate color when two colors of the same luminance are alternated at high frequencies~\cite{Lange1958ccff}. The critical color fusion frequency (CCFF), the frequency at which humans cannot perceive color vibrations, is approximately 25 Hz, which is about half of the critical flicker fusion frequency (CFF) at which luminance flicker becomes imperceptible~\cite{Mankowska2021CFFreview}. This imperceptible color vibration can be used without the need for dedicated high-refresh-rate displays since the refresh rates of commercially available LCD monitors and projectors exceed 60 Hz. Researchers are investigating the application of this principle to embed invisible information by rapidly altering the colors of specific areas of an image displayed on a screen, such as embedding invisible 2D QR codes that can be read by a camera~\cite{Abe2017-bu, Abe2020-md}.

Several studies have been conducted on efficient search methods for imperceptible color vibration pairs that can be detected by the color sensor~\cite{Abe2017-bu, Hattori2022-em}.
Recently, Hattori~\etal~\cite{hattori2024} further improved the efficiency of color search by using MacAdam ellipses~\cite{MacAdam1942-sv} as the basis for perceptual thresholds. They conducted user experiments to investigate the threshold for perceiving flicker by varying the amplitude of the color vibration along the long diameter of the MacAdam ellipse. In this paper, we extend their perceptual search for color vibration pairs to the application of gaze control.

\subsection{Color Vibration and Bottom-up Attention}\label{sec:related-attention}
Color vibrations are difficult to consciously perceive beyond CCFF.
However, while the ventral occipital (VO) cortex, which is involved in advanced color processing and visual cognition, does not perceive color vibrations, the earlier stages of the visual cortex V1 to V4 perceive color vibrations~\cite{Jiang2007}.
In light of this finding, it is hypothesized that since bottom-up attention originates from primitive visual stimuli in V1 to V4, although the color vibration is not perceived until they reach higher-order information processing areas in the brain, they may attract bottom-up attention in the visual cortex and evoke visual attention.
This hypothesis supports the underlying idea of this research.
Although accurate measurements of brain activity are needed to test this hypothesis in detail and accurately, this study can be positioned as a first pilot test of this hypothesis.

\section{Evaluation of Intermediate Perception in Color Vibration}\label{sec:eval-thresh}
We hypothesize that there is an intermediate state in the perception of color vibrations in which the object appears to be subtly different from a solid color, but without a clearly perceptible flicker.
Before considering gaze guidance using this intermediate state, this study first investigated whether such intermediate states can occur and, if so, in what range of amplitudes they fall.

Hattori~\etal~\cite{hattori2024} selected color vibration pairs based on MacAdam ellipses as a perceptual criterion and confirmed the threshold of color vibration amplitude at which flicker is not detectable by the human eye, using the major axis of the ellipses as a reference. 
This study further subdivides this color vibration detection threshold and investigates in detail, through user experiments, the vibration amplitude threshold for color pairs that produce the intermediate state.

\subsection{Selection of Perceptual Color Vibration Pairs}
\global\long\def\Matrix#1{\mathbf{\MakeUppercase{#1}}}
\global\long\def\Vector#1{\mathbf{\MakeLowercase{#1}}}
\global\long\def\Transpose{\mathrm{T}}

% \begin{figure}
%   \includegraphics[width=0.5\textwidth]{images/macadam_params.jpg}
%   \caption{(a) MacAdam ellipses on the $xy$ chromaticity diagram. (b) Selection method of color vibration pairs used in this study.}
%   \label{fig:macadam_params}
% \end{figure}

To select suitable color vibration pairs for our experiments, we used MacAdam ellipses to account for the non-uniformity in human color perception.

Figure~\ref{fig:teaser}
(a) shows the distribution of MacAdam ellipses on the $xy$ chromaticity diagram. Each ellipse represents a region where colors are indistinguishable from a center reference color when placed side by side, as determined experimentally.
By using MacAdam ellipses, we can select color vibration pairs that take into account the varying sensitivity of the human visual system to different colors. For example, the MacAdam ellipses are larger in the green region than in the blue region, indicating that small color differences in green are less perceptible than those in blue within the $xy$ chromaticity space.

Each MacAdam ellipse $\EllipseIdx=\{\CenterIdx, \RotationIdx, \MajorIdx, \MinorIdx\}~(\IndexE=1\cdots 25)$ is defined at 25 points on $xy$ chromaticity diagram by their center $\CenterIdx=[\CenterXIdx, \CenterYIdx]$, rotation angle $\RotationIdx$, and the lengths $\MajorIdx$ and $\MinorIdx$ of long and short diameter, respectively. We choose color pairs $\{\PairPIdx(r), \PairMIdx(r)\}$ multiplied by ratio $r$ along the long diameter $\MajorIdx$ of these ellipses as color vibration pairs (Fig.~\ref{fig:teaser} b), which denoted as
\begin{equation}
    \PairPMIdx(r)=[\CenterXIdx\pm r\cdot \MajorIdx \sin\RotationIdx,\;  \CenterYIdx \pm r\cdot \MajorIdx \cos\RotationIdx].
\end{equation}
This allows for the selection of color vibration pairs while considering the non-uniformity of human color perception. 

$xy$ chromaticity diagram is calculated by normalizing the luminance $0\leq Y\leq 1$ of the CIEXYZ color space. Therefore, when displaying colors based on the selected color pairs, it is necessary to complement the luminance. As luminance $Y$ approaches 0, the color pairs approach black, and the colors become nearly invisible. In contrast, as $Y$ approaches 1, the brightness of the color pairs exceeds the sRGB range.
Therefore, we set $Y=0.4$ and converted $xyY$ to XYZ as
\begin{equation}
    X=xY/y, \; Z=(1-x-y)Y/y.
\end{equation}
Then, color pairs selected in CIEXYZ are converted to the sRGB color system for display:
\begin{equation}
\begin{bmatrix}
R_{\Linear} \\
G_{\Linear} \\
B_{\Linear}
\end{bmatrix}
=
\begin{bmatrix}
3.2406 & -1.5372 & -0.4986 \\
-0.9689 & 1.8758 & 0.0415 \\
0.0557 & -0.2040 & 1.0570
\end{bmatrix}
\begin{bmatrix}
X \\
Y \\
Z
\end{bmatrix} 
\end{equation}
The following gamma transformations were then applied to each channel $C\in\{R, G, B\}$:
% \begin{eqnarray}\label{eq:gamma-correct}
% C_{\srgb} &=& \min(\max(\gamma(C_{\Linear}), 0), 1), \\
% \gamma(C_{\Linear}) &=& \begin{cases}
% 1.055 \cdot C_{\Linear}^{1/2.4} - 0.055 & \text{if $c\geq 0.0031308$,} \\
% 12.92 & \text{otherwise}
% \end{cases}
% \end{eqnarray}
\begin{eqnarray}\label{eq:gamma-correct}
C_{\srgb} &=& \gamma(C_{\Linear}) \\
&=& \begin{cases}
1.055 \cdot C_{\Linear}^{1/2.4} - 0.055 & \text{if $c\geq 0.0031308$,} \\
12.92 \cdot C_{\Linear} & \text{otherwise}
\end{cases}
\end{eqnarray}

For implementation, we use the color science library in Python~\footnote{Colour 0.4.4 by Colour Developers, https://zenodo.org/records/10396329}.
The CIE 1931 2$^\circ$ observer function under D65 illumination is used for color conversion.

Hattori~\etal~\cite{hattori2024} previously selected pairs of color vibrations by varying $r$ according to the above criteria and presented them to the participants.
They derived a threshold of $r=24.4$ at which 50~\% of the participants perceived color vibration, and considered values below this threshold as imperceptible color vibration and applied it to the detection of color vibration by cameras.
In contrast, we further subdivide the perception of color vibration and evaluate the range of $r$ that results in a state of ``different from a solid color but not clearly flickering''.

\subsection{Experiment Setup}~\label{sec:exp1-setup}

\subsubsection{Participants}
17 participants (13 males, 4 females; age range 23--45 years, mean age 30.3 years) participated in the study. 9 participants used corrective lenses (glasses or contact lenses). The study was approved by the Institutional Review Board at [name of institution, omitted for double-blind review].

\subsubsection{Generation of Color Vibration Images}

In this experiment, we varied three parameters: the amplitude ratio $r$, the diameter $d$ of the image circle, and the display position of the image. We presented various combinations of color vibrations to participants and measured human perceptual thresholds based on their responses.

Participants were asked to choose one of the following three perceptual states:
\begin{enumerate}[leftmargin=*]
\item \textbf{Indistinguishable from a solid color}: The image appears as a uniform color with no noticeable differences.
\item \textbf{Different from a solid color, but not clearly flickering}: The image appears slightly different from a solid color, but there is no noticeable flickering.
\item \textbf{Clearly flickering}: The image has noticeable flickering.
\end{enumerate}
We selected the point $(x, y) = (0.305, 0.323)$ from the MacAdam ellipses in the $xy$ chromaticity space as the base color for generating color vibration pairs. The luminance was set to $Y = 0.4$. This color is near the center of the $xy$ chromaticity diagram and ensures that the color pairs remain within the sRGB gamut even as $r$ increases. In the sRGB color space, this corresponds to $(R, G, B) = (163, 168, 173)$. We generated color pairs with $r$ values ranging from 0 to 50 in increments of 5, resulting in 11 sets of color vibration pairs.

Note that, in this paper we focus only on vibrating grayscale colors, since individual perception of color vibration varies with hue and saturation. The investigation of the parameter $r$ for any color is left for future work.

\begin{figure}[t]
  \centering
  \includegraphics[width=\columnwidth]{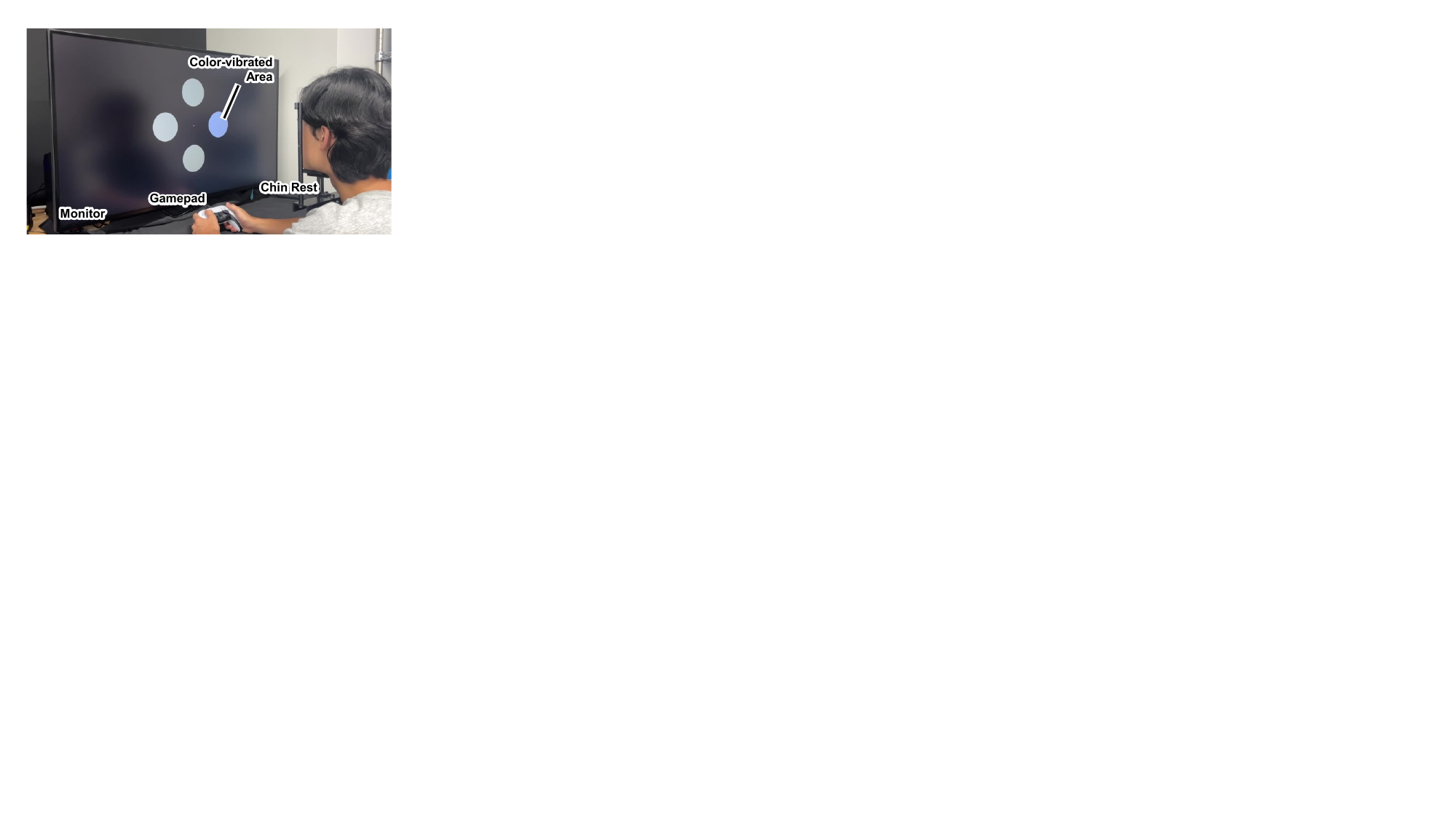}
  \caption{Experimental setup for Sec.~\ref{sec:eval-thresh}.}
  \label{fig:exp1-setup}
\end{figure}

\begin{figure}[t]
  \centering
  \includegraphics[width=\columnwidth]{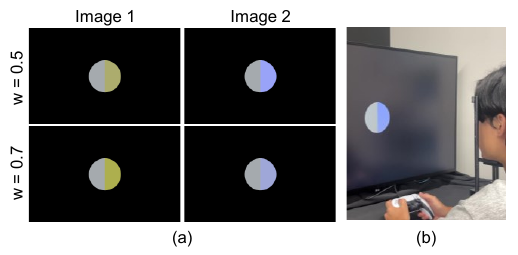}
  \caption{(a) Image pairs to be presented during color vibration adjustment ($r=50$) for each participant. The image pair at $w=0.7$ has a stronger yellow saturation than the image pair at $w=0.5$. (b) Color vibration adjustment. The participant adjusts the weight $w$ so that the color appears perceptually the same as the gray of the left half.}
  \label{fig:exp-color-fitting}
\end{figure}

\begin{figure*}[t]
  \centering
  \includegraphics[width=\linewidth]{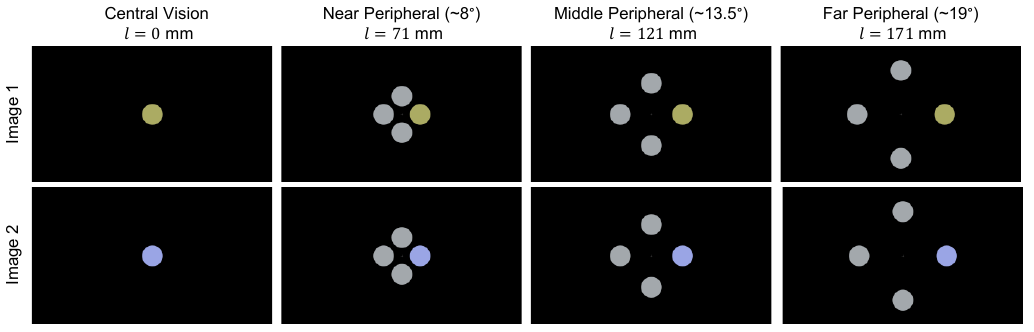}
  \caption{An example of the color vibration image pairs presented in the experiment of Sec.~\ref{sec:eval-thresh}. In the central visual field condition, we applied color vibration at a specified $r$ and $d$ to a single circle. In the peripheral visual field condition, we applied color vibration at the specified $r$ and $d$ to one of four circles. Participants were asked to indicate whether they perceived the color vibration.}
  \label{fig:exp1-display-imgs}
\end{figure*}

\begin{figure*}[t]
  \centering
  \includegraphics[width=\linewidth]{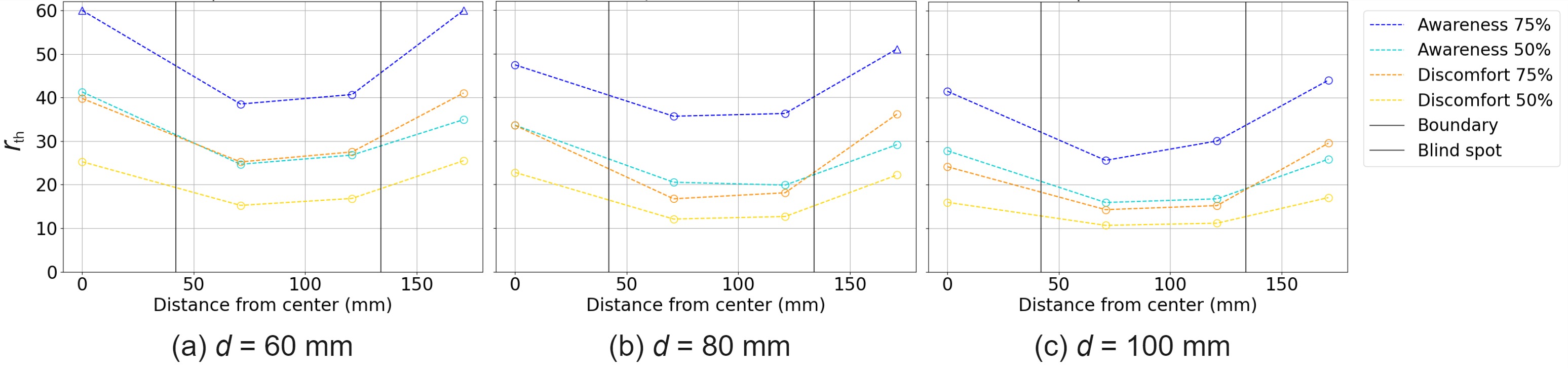}
  \caption{Experimental results showing the thresholds $r_{\mathrm{th}}$ for the Awareness condition (participants perceive the color as ``different from a solid color'') at 50\% and 75\% probability, and the thresholds for the Discomfort condition (participants perceive an ``obvious flicker''), for each display position of the image circle. The image circle sizes are $d = 60~\mathrm{mm}, 80~\mathrm{mm}, 100~\mathrm{mm}$. Note that Boundary represents the boundary between central and peripheral vision, and Blind spot corresponds to the blind spot in human vision.}
  \label{fig:result-thresh}
  \vspace{-3mm}
\end{figure*}

\subsubsection{Apparatus}
Participants were seated with their chins on a rest to stabilize head position, ensuring their eyes were level with the center of a 42.5-inch LCD display (43UN700-BAJP, LG Electronics) positioned 500 mm away. The display was calibrated to the sRGB color space using a monitor calibration tool (Spyder X Elite, Datacolor) and set to a luminance of 166 cd/m\textsuperscript{2}.

% \subsubsection{Experiment Condition and Procedure}
% Figure~\ref{fig:exp1-setup} shows the setup of the experiment.
% 17 participants (13 males, 4 females, aged 23-45, mean age 30.3 years) participated in this experiment.
% Nine participants had their vision corrected with glasses or contact lenses.
% Participants were presented with color vibration pairs on a 42.5-inch LCD display (43UN700-BAJP, LG Electronics).
% 我々は、このディスプレイをモニタキャリブレーションツール（Spyder X Elite, datacolor）を用いてsRGBの色域に色校正を行い、ディスプレイ輝度を166 cd/m2以下に設定した。
% The participants rested their chins on a chin rest placed in front of the display, with the height of their eyes set at the center of the monitor and the distance between the participant and the display adjusted to 500 mm.

\subsubsection{Experimental Conditions and Procedure}~\label{sec:exp1-exp-condition}
Figure~\ref{fig:exp1-setup} shows the experimental setup.
Prior to the main experiment, we conducted a color adaptation procedure to account for individual differences in color perception. Participants were shown an image circle (diameter 120 mm) centered on the display, which was vertically split: the left half showed a solid color without vibration, and the right half showed a color-vibrating image at a specific $r$ value. Participants adjusted the parameter $w$ -- which determined the weight between the two colors in the vibrating pair -- using the left and right buttons of a gamepad in increments of 0.02, aiming to make the left and right halves appear identical (Fig.~\ref{fig:exp-color-fitting}).

Mathematically, the vibrating color pair was adjusted so that the ratio of the distances along the major axis of the MacAdam ellipse from the target color to the vibrating colors (yellowish side and bluish side) was $w : (1 - w)$, where $0 < w < 1$. This adjustment personalized the center of the color vibration to each participant's perception.

After each adaptation, an inverted stimulus image was displayed for 100 ms to reduce afterimage effects, where the image circle turned black and the surrounding background remained the display color. This fitting process was repeated five times with decreasing $r$ values ($r = 50, 40, 30, 20, 10$) to familiarize participants with the range of color vibrations and to help them understand the concept of "flicker."

Figure~\ref{fig:exp1-display-imgs} shows an example of the images presented in the main experiment. In the main experiment, color vibration images were displayed as circles with diameters $d$ set to 60 mm (small), 80 mm (medium), or 100 mm (large). The positions of the image circles were determined by the distance $l$ from the center of the display to the center of the circle, which was set to $l = 0$, 71, 121, or 171 mm. These distances correspond to central vision (0° visual angle), near peripheral vision ($\sim$ 8$^\circ$), middle peripheral vision ($\sim$ 13.5$^\circ$), and far peripheral vision ($\sim$ 19$^\circ$), respectively. For positions other than central vision, four image circles were symmetrically arranged around the center to eliminate directional bias.

Participants were presented with a random sequence of the 132 combinations ($11 \times 3 \times 4$) of $r$, $d$, and $l$. On each trial, they used a gamepad (DualSense Wireless Controller, Sony Interactive Entertainment) to select one of the three perceptual states described above. When multiple image circles were displayed (for peripheral vision conditions), participants also indicated which circle they perceived as vibrating.
An inverted stimulus was displayed after each response to prevent afterimages. Participants were given short breaks after every 10 trials to reduce fatigue.

\subsection{Result and Discussion}
We calculated the proportion of participants who reported each perceptual state at different $r$ values for each condition. By fitting sigmoid functions to the data, we estimated the thresholds $r_{\mathrm{th}}$ at which participants had a 50~\% and 75~\% probability of perceiving the image as "different from a solid color" and "clearly flickering".
In the peripheral vision conditions, if a participant incorrectly identified which of the four image circles was vibrating, we assumed that they did not perceive an anomaly.

Figure~\ref{fig:result-thresh} shows the experiment result. 
The graph shows, from left to right, the $r$ threshold at which subjects perceive a color as different from a single color with 50\% and 75\% probability, and the $r$ threshold at which they perceive an obvious flicker, for each display position of the image circle when the size of the image circle $d=60\ \mathrm{mm}, 80\ \mathrm{mm}, 100\ \mathrm{mm}$.

\subsubsection{Effect of Color Vibration Size}
Figure~\ref{fig:result-thresh} shows that as the image circle diameter $d$ increases, the threshold $r_{\mathrm{th}}$ decreases. 

Specifically, for $d = 60$, 80, and 100 mm, the lowest thresholds at which participants perceived the image as ``different from a solid color'' with 50~\% probability were $r_{\mathrm{th}} = 25.22$, 16.73, and 14.29 at $l = 71$ mm, and with 75~\% probability were $r_\mathrm{th}=38.50$, $35.65$, $25.61$ at $l$ = 71 mm, respectively.
The thresholds for perceiving ``obvious flicker'' with 50~\% probability were lower at $r_{\mathrm{th}} = 15.25$, 12.08, and 10.68, respectively, and with 75~\% probability were $r_\mathrm{th}=25.22$, $16.73$, $14.29$, respectively.
This suggests that larger color vibration areas make it easier for participants to perceive color vibrations even at smaller $r$ values.

\subsubsection{Effect of Display Position}
From Fig.~\ref{fig:result-thresh}, the thresholds varied with display position. Perception of color vibration was less sensitive in central vision ($l = 0$) compared to near and middle peripheral vision ($l = 71$, 121 mm). However, in the central vision condition, only one circle was displayed, eliminating the possibility of color comparison with adjacent stimuli, which may have contributed to the higher thresholds observed.

In far peripheral vision ($l = 171$ mm), sensitivity to color vibration decreased compared to near and middle peripheral vision. This indicates that the sensitivity to color vibration depends on the position within the visual field, likely due to the visual system's characteristics, such as reduced spatial and temporal resolution in peripheral areas.

\subsubsection{Intermediate Perceptual State of Color Vibrations}
The observed difference between the thresholds for "different from a solid color" and "clearly flickering" supports the existence of an intermediate perceptual state. If such a state did not exist, these thresholds would be nearly identical under the same conditions.
This indicates that the subject first perceives the color difference from the monochromatic color, and then there is a process of flicker perception. 

These results show that the perception of color vibration is influenced by factors beyond the amplitude ratio $r$, including image size and display position. 
By choosing appropriate $r$ values between the "different from a solid color" and "clearly flickering" thresholds for each condition, we can design effective gaze guidance that avoids causing noticeable flicker.

\section{Evaluation of Color Vibration Vision Guidance}\label{sec:eval-guidance}
\begin{figure}[t]
  \centering
  \includegraphics[width=\columnwidth]{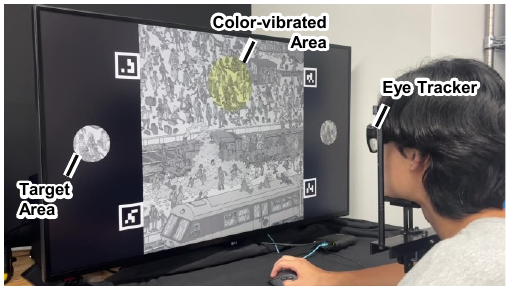}
  \caption{Experimental setup for Sec.~\ref{sec:eval-guidance}.}
  \label{fig:exp2-setup}
  \vspace{-3mm}
\end{figure}

Next, based on the findings in Sec.~\ref{sec:eval-thresh}, we conducted an experiment to investigate whether applying color vibration to a region of interest (ROI) in an image could attract users' attention without altering the overall impression of the image. Figure~\ref{fig:exp2-setup} shows the experimental setup.
 The experimental protocol was approved by the [Omitted for double-blind review] Research Ethics Committee.

\subsection{Experiment Setup}
\subsubsection{Participants}
29 participants (20 males and 9 females; age range 23-50 years, mean age 30.7 years), none of whom had participated in the previous experiment, participated in this study. 6 participants corrected their vision with glasses and 12 used contact lenses.

\subsubsection{Apparatus}
Participants were seated with their chins resting on a chin rest placed in front of the same 42.5-inch LCD display used in Sec.~\ref{sec:exp1-setup}. 
The display settings, viewing distance (500 mm), and ambient lighting conditions were the same as in the previous experiment. For 23 participants who did not wear glasses, we tracked and recorded their gaze using eyeglass-mounted eye trackers (Tobii Pro Glasses 3). To map gaze data to display coordinates, ArUco markers of known dimensions were displayed at the four corners of the content images, allowing homography transformation using the eye tracker's cameras.

\begin{figure*}[t]
  \centering
  \includegraphics[width=\textwidth]{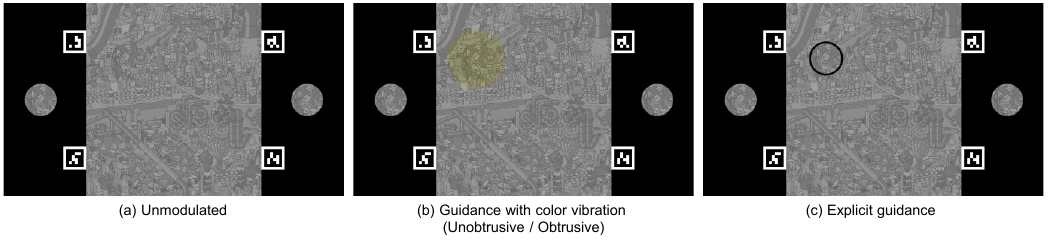}
  \caption{Examples of images presented in Sec.~\ref{sec:eval-guidance}. (a) The unmodulated image,~\ie, the original image. (b) Images with color vibration applied to the ROI. (c) The "explicit" VG image with the ROI circled.}
  \label{fig:exp2-1-samples}
\end{figure*}

\subsubsection{Stimuli Preparation}
Figure~\ref{fig:exp2-1-samples} shows an example of the presentation image used in this experiment.
For experiment images, we used images from "Pocket Edition NEW Where's Wally!" by Martin Handford, a picture book in which readers search for a specific character among many others. Note that, in this experiment, the character that the participants have to find is not necessarily Wally, as they are asked to look at objects in various positions on the screen.

6 images were selected and scanned at 600 dpi using a scanner (ScanSnap iX1300, PFU). The images were cropped to 1200 px $\times$ 1200 px to adjust the difficulty level to ensure that participants could locate the target character within the 30 second display limit even in the unmodulated condition.
To facilitate the application of color vibration pairs to each pixel, images were converted to grayscale and pixel values were remapped from the original range [0, 255] to [60, 196]. The ROI was defined as a circular area 44 mm in diameter sufficient to enclose a single character (not necessarily Wally) and approximating the effective area of central vision (a circle 39 mm in diameter at 500 mm viewing distance). The positions of the ROIs were varied across images to avoid location bias.

The area of color vibration was defined as an 80 mm diameter circle encompassing the ROI. We applied four variations of visual guidance to the images:
\begin{enumerate}[leftmargin=*]
    \item \textbf{Unmodified}: Original image without any modification.
    \item \textbf{Unobtrusive color vibration}: ROI modulated with subtle color vibration.
    \item \textbf{Obtrusive Color Vibration}: ROI modulated with more pronounced color vibration.
    \item \textbf{Explicit guidance}: ROI marked with a visible circle.
\end{enumerate}

% \subsubsection{色振動画像の生成}

\subsubsection{Determination of Color Vibration Parameters}
The color vibration pairs were generated using the threshold values $r_{\mathrm{th}}$ corresponding to the diameter $d$ of the vibrating area and the distance $l$ from the center of the display, as obtained in the previous experiment. Specifically, we used the values at which participants perceived the color vibration with 75~\% probability for $d = 80$ mm. We chose the 75~\% threshold to ensure that the cue was effective in attracting users' attention.

For the unobtrusive color vibration, we used the $r_{\mathrm{th}}$ corresponding to the "Awareness" condition, while for the obtrusive color vibration, we used the $r_{\mathrm{th}}$ corresponding to the "Discomfort" condition. If the ROI was in the central visual field (within 5 degrees of visual angle, $l = 0$ mm), we used the corresponding $r_{\mathrm{th}}$. For ROIs in the peripheral visual field (approximately 8$^\circ$ to 19$^\circ$ of visual angle), we linearly interpolated the $r_{\mathrm{th}}$ values based on the results for $l = 71$, 121, and 171 mm. We avoided placing ROIs outside of these ranges due to uncertainties in the applicability of linear interpolation.

Because the parameter $r$ varied with distance from the center of the display, we adjusted the color-fitting parameter $w$ accordingly for each participant, using linear interpolation based on the fitting results from the previous experiment.

\subsubsection{Experiment Procedure}
Participants provided informed consent and underwent the same color fitting process as in Sec.~\ref{sec:exp1-exp-condition}. Definitions of terms used in the Likert scale questionnaire were then explained. The eye tracker was calibrated before the experiment began.

During the experiment, participants rested their chin on the chin rest to maintain a constant viewing position. They were instructed to keep their gaze on the display and to use a mouse to select the target area within the content images. Reference images of the search target were displayed on either side of the content image for participants to refer to as needed.

Each participant viewed 6 sets of images, each set containing one of the 4 visual guidance conditions described above. The order of the image sets and the images within each set were randomized.

For each trial, the order was as follows:
\begin{enumerate}[leftmargin=*]
    \item \textbf{Fixation screen}: A black screen with a white cross in the center was displayed, and participants were instructed to fixate on the cross.
    \item \textbf{Target presentation}: An image showing only the ROI (the target character) centered on the screen was presented to familiarize participants with the search target.
    \item \textbf{Search task}: The content he assigned visual guidance was displayed. Participants searched for the target character and indicated their selection with a mouse click.
    \item \textbf{Questionnaire}: After each image, participants rated the following on a 7-point Likert scale, following Sutton et al.~\cite{Sutton2022}:
    \begin{itemize}[leftmargin=*]
        \item \textbf{Naturalness}: The extent to which the image appeared unprocessed (1: very unnatural -- 7: very natural).
        \item \textbf{Obtrusiveness}: The degree to which the image stood out undesirably (1: not at all obtrusive -- 7: very obtrusive).
    \end{itemize}
\end{enumerate}
After every 6 images, participants took a short break and looked at a black screen to rest their eyes.

\subsection{Results}
\begin{figure}[t]
  \centering
  \includegraphics[width=\columnwidth]{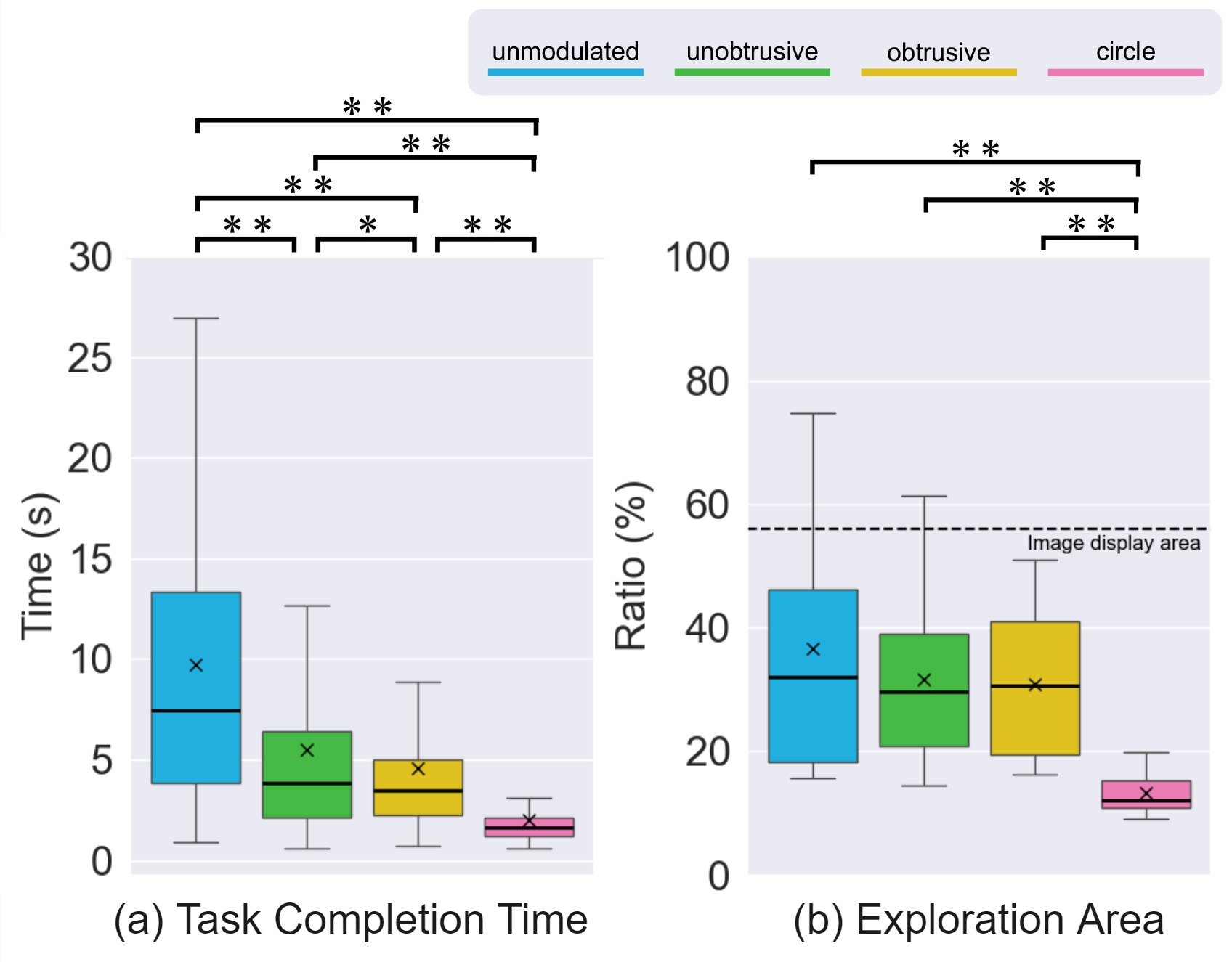}
  \caption{Results of (a) task completion time and (b) ratio of the area explored relative to the entire screen}
  \label{fig:result_task_completion}
  \vspace{-3mm}
\end{figure}

% \begin{figure}[t]
%   \centering
%   \includegraphics[width=\columnwidth]{images/5_heatmap_overlied_with_colorbar.png}
%   \caption{Heatmap of gaze under unobtrusive color conditions}
%   \label{fig:heatmap_unobtrusive}
% \end{figure}

\begin{figure}[t]
  \centering
  \includegraphics[width=\columnwidth]{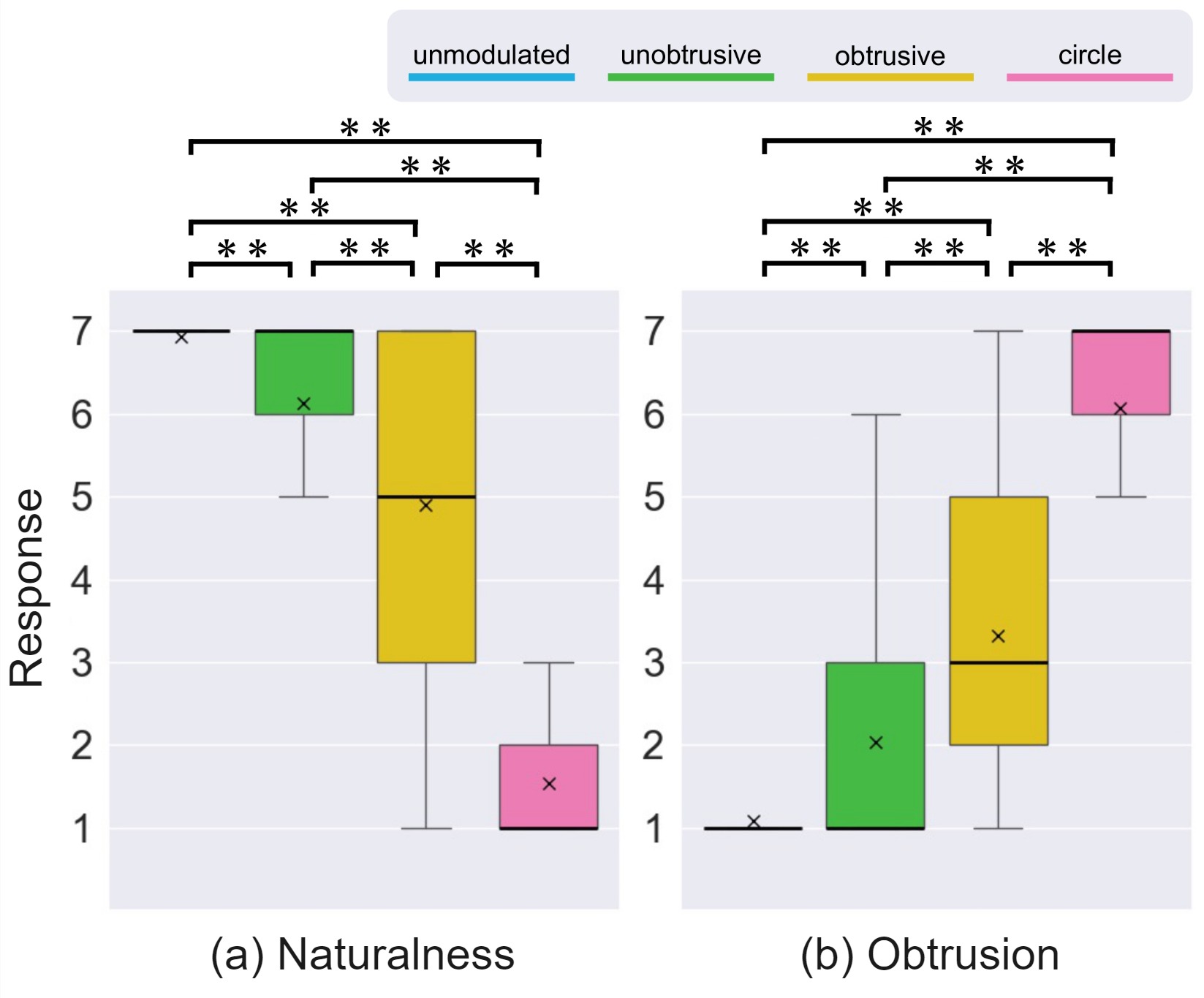}
  \caption{Results of a user questionnaire on the four display conditions.}
  \label{fig:result_questionnaire}
  \vspace{-3mm}
\end{figure}

\begin{figure*}[t]
  \centering
  \includegraphics[width=\linewidth]{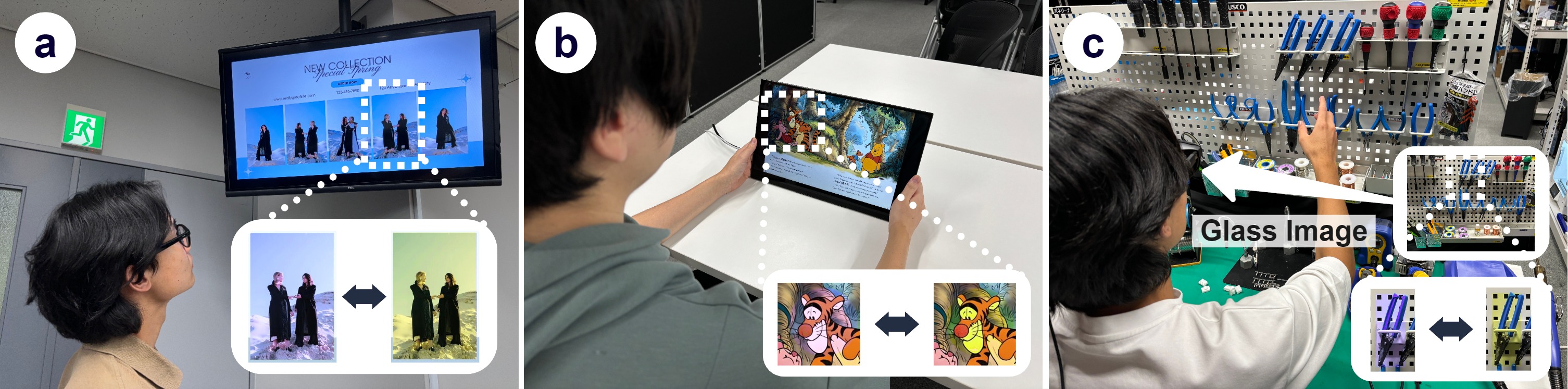}
  \caption{Application scenarios utilizing unobtrusive color vibration for visual guidance: (a) Advertising: directing user attention to specific products or information within an advertisement without compromising design aesthetics; (b) Picture book: guiding readers' attention to important characters or key elements in a digital book while maintaining immersion; (c) Task assistance: helping users wearing AR glasses to identify necessary tools or parts during tasks by unobtrusively highlighting them, thereby improving work efficiency and reducing errors.}
  \label{fig:application}
  \vspace{-3mm}
\end{figure*}

We evaluated the experiment from three perspectives: (1) task completion time (time to click on the correct search area), (2) proportion of the image explored by the gaze before task completion, and (3) scores from the user questionnaire.

\subsubsection{Task Completion Time}
Time was recorded based on participants' responses. Correct responses were assigned their respective completion times, while incorrect responses or timeouts (exceeding the 30-second limit) were assigned a 30-second time. 

Figure~\ref{fig:result_task_completion} (a)
shows the task completion times for the four display conditions. As a result of conducting the Friedman test, the main effect of the display conditions on the evaluation score was significant ($\chi^2$ = 252.730, \textit{p} < .001).
Since the main effect was significant, the Wilcoxon signed-rank test was repeated under the Holm method as a sub-test.
As a result, the task completion time of unmodulated condition was significantly higher than the unobtrusive condition (\textit{p} < .001, Cohen's \textit{r} = 0.425), the obtrusive condition (\textit{p} < .001, Cohen's \textit{r} = 0.598), and the circle condition (\textit{p} < .001, Cohen's \textit{r} = 0.835).
The unobtrusive condition was also significantly higher than the obtrusive condition (\textit{p} = .0011, Cohen's \textit{r} = 0.192) and the circle condition (\textit{p} < .001, Cohen's \textit{r} = 0.731).
The obtrusive condition was also significantly higher than the circle condition (\textit{p} < .001, Cohen's \textit{r} = 0.748).

\subsubsection{Proportion of Explored Area}
We calculated the proportion of the image explored by the gaze using the eye-tracking data. For each gaze point, we considered the area within the central visual field (within 5 degrees of visual angle) as the exploration area. The total exploration area was divided by the total image area to obtain the proportion. 
The eye coordinate information was converted to display coordinate system coordinates by performing a homographic transformation using the coordinates of the ArUco marker detected by the eye tracker camera.

Figure~\ref{fig:result_task_completion} (b) shows the results. As a result of conducting the Friedman test, the main effect of the display conditions on the evaluation score was significant ($\chi^2$ = 38.018, \textit{p} < .001).
Since the main effect was significant, the Wilcoxon signed-rank test was repeated under the Holm method as a sub-test.
As a result, the ratio of exploration area of unmodulated condition was significantly higher than  the circle condition (\textit{p} < .001, Cohen's \textit{r} = 0.876), the unobtrusive condition was significantly higher than  the circle condition (\textit{p} < .001, Cohen's \textit{r} = 0.876), and the obtrusive condition was significantly higher than the circle condition (\textit{p} < .001, Cohen's \textit{r} = 0.869).
Whereas no significant differences were observed between the unmodulated and unobtrusive conditions (\textit{p} = .248, Cohen's \textit{r} = 0.253), between the unmodulated and obtrusive conditions (\textit{p} = .129, Cohen's \textit{r} = 0.329), and between the unobtrusive and obtrusive conditions (\textit{p} = .924, Cohen's \textit{r} = 0.024).

\subsubsection{User Questionnaire}
Figure~\ref{fig:result_questionnaire} shows the questionnaire results for naturalness and obtrusiveness.

As a result of conducting the Friedman test, the main effect of the display conditions on the evaluation score was significant for all categories of questions (Naturalness: $\chi^2$ = 433.243, \textit{p} < .001; Obtrusion: $\chi^2$ = 390.673, \textit{p} < .001).
Since the main effect was significant, the Wilcoxon signed-rank test was repeated under the Holm method as a sub-test.
The results are shown below:
\begin{itemize}[leftmargin=*]
\setlength{\parskip}{0.1cm}
    \item \textbf{Naturalness}: the score of unmodulated condition was significantly higher than the unobtrusive condition (\textit{p} < .001, Cohen's \textit{r} = 0.863), the obtrusive condition (\textit{p} < .001, Cohen's \textit{r} = 0.861), and the unobtrusive condition (\textit{p} < .001, Cohen's \textit{r} = 0.867). The score of the unobtrusive condition was also significantly higher than the obtrusive condition (\textit{p} < .001, Cohen's \textit{r} = 0.850) and the circle condition (\textit{p} < .001, Cohen's \textit{r} = 0.866). The score of the obtrusive condition was also significantly higher than the circle condition (\textit{p} < .001, Cohen's \textit{r} = 0.866).
    \item \textbf{Obtrusion}: the score of unmodulated condition was significantly lower than the unobtrusive condition (\textit{p} < .001, Cohen's \textit{r} = 0.861), the obtrusive condition (\textit{p} < .001, Cohen's \textit{r} = 0.863), and the unobtrusive condition (\textit{p} < .001, Cohen's \textit{r} = 0.866). The score of the unobtrusive condition was also significantly lower than the obtrusive condition (\textit{p} < .001, Cohen's \textit{r} = 0.849) and the circle condition (\textit{p} < .001, Cohen's \textit{r} = 0.864). The score of the obtrusive condition was also significantly lower than the circle condition (\textit{p} < .001, Cohen's \textit{r} = 0.851).
\end{itemize}

\subsection{Discussion}
The task completion time result showed that there were significant differences between all four display conditions.
This result indicates that unobtrusive gaze guidance using color vibration effectively directs users' attention and improves task performance without significantly altering the overall scene. While its effectiveness is slightly lower than that of obtrusive color vibration and explicit guidance, it offers sufficient performance as a subtle gaze guidance method, avoiding excessive interference with the user experience.

In terms of the proportion of explored area, significant differences were observed only between the explicit guidance condition and the other three conditions. This suggests that gaze guidance using color vibration allows users to naturally explore the entire scene, unlike explicit guidance, which strongly directs gaze to a specific area and may limit exploration. With color vibration, users can broadly view the scene while still being subtly guided toward important information.

The questionnaire results show that unobtrusive gaze guidance was perceived as more natural and less obtrusive compared to obtrusive color vibration and explicit guidance. Although rated slightly less natural than the unmodulated condition, unobtrusive gaze guidance successfully attracted users' attention without significantly affecting the scene context.

In the user questionnaire results, significant differences were observed among all display conditions in the evaluations of naturalness and obtrusiveness. Specifically, unobtrusive gaze guidance using color vibration was found to have significantly higher naturalness compared to obtrusive color vibration and explicit guidance. This is because unobtrusive gaze guidance can attract the user's attention without significantly altering the scene's context. On the other hand, since the obtrusiveness rating was higher than that of the unmodulated condition, it indicates that the unobtrusive gaze guidance is effectively capturing the user's attention.

In terms of obtrusiveness, the unobtrusive gaze guidance showed less negative impact on the scene's context compared to explicit guidance. Explicit gaze guidance may force the user's attention toward a specific area, potentially impairing overall scene comprehension and immersion. In contrast, unobtrusive gaze guidance can promote attention to necessary information while respecting the user's natural eye movements.

From these results, it is suggested that unobtrusive gaze guidance using color vibration is an effective method that contributes to improving task completion speed and naturally supports the user's exploratory behavior. Because it can enhance access to necessary information without diminishing the user experience, practical applications are anticipated.

\section{Application}
We explored application scenarios for ChromaGazer proposed in this study and evaluated its effectiveness. Specifically, we propose the potential of using unobtrusive color vibration for visual guidance in three applications: advertising, picture books, and task assistance. Figure~\ref{fig:application} illustrates these applications.

In the advertising application (Fig.~\ref{fig:application}a), we selected a region of interest (ROI) within an advertisement image and generated a pair of images with unobtrusive color vibration for that region. By displaying this image, it becomes possible to direct the user's attention to specific products or information without compromising the design or aesthetic appeal of the advertisement. Unlike conventional pop-ups or explicit banners, unobtrusive visual guidance can enhance advertising effectiveness without disrupting the user experience.

In the picture book application (Fig.~\ref{fig:application}b), we unobtrusively guided the reader's attention to specific regions within the images of a digital book, synchronized with the progression of the story. This allows readers to naturally focus on important characters or key elements of the narrative while maintaining immersion in the world of the picture book. Unobtrusive visual guidance supports readers in smoothly following the story's development, contributing to improved readability and comprehension.

In the task assistance application (Fig.~\ref{fig:application}c), we envisioned a scenario where a user wearing AR glasses performs work using tools. By detecting in real time the regions of necessary tools or parts during the task and applying unobtrusive color vibration to those areas, the user's attention can be naturally guided. This enables the user to quickly identify the next tool to use or the component to focus on without obstructing the view with overt guidance information. This approach is expected to lead to improvements in work efficiency and reduction of errors.

Through these application scenarios, the potential of ChromaGazer to achieve effective visual guidance while maintaining user immersion and a natural experience was demonstrated. By using unobtrusive color vibration, it is possible to direct the user's attention to necessary information without significantly altering the context of the scene.

\section{Limitation and Future Work}
While this study is the first to demonstrate gaze guidance using color vibration, several areas require further investigation in order to generalize the perceptual parameters across different situations. We discuss the current limitations and potential avenues for future work.

\paragraph{Relation between color vibration amplitude and hue/brightness} 
We focus exclusively on vibrating grayscale colors because the perception of color vibration varies according to hue and saturation and we wanted to see if color vibration is effective in VG by controlling variables. Having verified the effectiveness of color vibration on VG in this paper, investigating the effect of color differences on amplitude parameters is the most important future work.

Moreover, our experiments assumed that the amplitude of chromaticity for a given hue is proportional to the major axis of the MacAdam ellipse. However, the perception of chromaticity may vary with hue and lightness. Future studies could model the amplitude $r$ as a function of hue and lightness through comprehensive experiments.

\paragraph{Generalized saliency model for color vibration} The images used in the current experiment are limited, and it is necessary to verify the effectiveness of the method for various scenes and objects in natural images. If an appropriate saliency model for color vibration is provided, the VG effect of color vibration will be evaluated more generally. However, at present, no appropriate model for saliency control of color vibration has been proposed, and this is positioned as a future research direction.

\paragraph{Neuroscientific exploration of the intermediate state of Color Vibration} As introduced in Sec.~\ref{sec:related-attention}, the intermediate state of color vibration explored in this study was hypothesized based on neuroscientific findings. While this paper focused on user subjective evaluation and gaze analysis, estimating the intermediate state based on brain measurements such as fMRI is expected to deepen this method further.

\paragraph{Addressing individual differences} The perceptual characteristics of color vibration depend on the individual, and experiments with larger numbers of participants are needed. In particular, since the participants in our experiment were limited to a young age group, it is necessary to demonstrate the effect across a wider age range. In addition, the calibration and personalization of individual color vibration perception, including color vision deficiency~\cite{langlotz2018chromaglass}, is considered a future research direction.

\paragraph{Experiments in different display environments} The current experiment demonstrated saliency control using color vibration through measurements on a color-calibrated monitor under natural light conditions. However, further research is needed on the relationship between monitor brightness, ambient light brightness, and color vibration perception parameters. In addition, research on applying gaze control to VR~\cite{Sitzmann2018saliency, Grogorick2020} and AR~\cite{Sutton2022, Sutton2024flicker} environments that modulate real-world saliency is also progressing. Achieving optimal color vibrancy modulation in these environments is positioned as a future research challenge.

\section{Conclusion}
We introduced ChromaGazer, a visual guidance technique leveraging color vibration to direct users' attention without perceptibly altering image appearance. Through user experiments, we determined thresholds for the intermediate perceptual state of color vibration, informing ChromaGazer's design. Evaluations with natural images demonstrated the method's effectiveness in guiding gaze while preserving naturalness compared to existing approaches. We also assessed the impact of color vibration amplitude on gaze guidance efficacy, emphasizing the need for optimal amplitude selection.
This study opens opportunities for color vibration-based gaze guidance techniques. Generalizing our method requires further investigation into the perceptual characteristics of color vibration and the development of advanced gaze guidance strategies in diverse environments. We hope this research stimulates further exploration toward more sophisticated visual guidance systems.

% In this paper, we introduced ChromaGazer, a VG technique that exploits the perceptual properties of color vibration to direct viewers' attention to desired regions without perceptually altering the appearance of the image. User experiments were conducted to determine the threshold for the intermediate perceptual state of color vibration, which informed the design of ChromaGazer. Evaluations using natural images demonstrated the effectiveness of our method in guiding the gaze while maintaining naturalness compared to existing approaches. The effect of color vibration amplitude on the effectiveness of gaze guidance was also evaluated, highlighting the importance of selecting an optimal amplitude.

% This work provides new avenues for color vibration-based gaze guidance techniques. Generalizing the method would require further investigation into the perceptual characteristics of color vibration and the development of more advanced gaze guidance techniques in different environments. We anticipate that this research will catalyze further exploration, ultimately creating more sophisticated VG systems.

\acknowledgments{
This study was supported by JST PRESTO Grant Number JPMJPR17J2 and JSPS KAKENHI Grant Number JP20H04222, JP23H04328, JP24KK0187, Japan.
}

\bibliographystyle{abbrv}

\raggedbottom
\bibliography{references}

\end{document}

%% file: macros.tex
\global\long\def\Matrix#1{\mathbf{\MakeUppercase{#1}}}

\global\long\def\Vector#1{\mathbf{\MakeLowercase{#1}}}

\global\long\def\Rotation{\Matrix R}

\global\long\def\Transpose{\mathrm{T}}

\global\long\def\IndexE{n}

\global\long\def\Element#1#2{#1_{#2}}

\global\long\def\Ellipse{\mathcal{E}}
\global\long\def\Center{\Vector{c}}
\global\long\def\Rotation{\theta}
\global\long\def\Major{a}
\global\long\def\Minor{b}

\global\long\def\EllipseIdx{\Element{\Ellipse}{\IndexE}}
\global\long\def\CenterIdx{\Element{\Center}{\IndexE}}
\global\long\def\RotationIdx{\Element{\Rotation}{\IndexE}}
\global\long\def\MajorIdx{\Element{\Major}{\IndexE}}
\global\long\def\MinorIdx{\Element{\Minor}{\IndexE}}

\global\long\def\CenterXIdx{\Element{c}{\IndexE x}}
\global\long\def\CenterYIdx{\Element{c}{\IndexE y}}

\global\long\def\PairP{\Vector{p}^{+}}
\global\long\def\PairM{\Vector{p}^{-}}
\global\long\def\PairPM{\Vector{p}^{\pm}}

\global\long\def\PairPIdx{\Element{\PairP}{\IndexE}}
\global\long\def\PairMIdx{\Element{\PairM}{\IndexE}}
\global\long\def\PairPMIdx{\Element{\PairPM}{\IndexE}}

\global\long\def\Linear{\mathrm{linear}}
\global\long\def\srgb{\mathrm{srgb}}